\documentclass[prb,superscriptaddress,twocolumn,showpacs]{revtex4}

\usepackage{graphicx}

\begin{document}
\newcommand{\icm}{\ensuremath{\mbox{cm}^{-1}}}
\newcommand{\SF}{(TMTSF)$_2$\-PF$_6$}
\newcommand{\TF}{(TMTTF)$_2$\-PF$_6$}
\newcommand{\asf}{(TMTTF)$_2$\-AsF$_6$}

\title{Deconfinement transition and dimensional crossover in the Bechgaard-Fabre salts:\\
pressure- and temperature-dependent optical investigations}

\author{A. Pashkin}
\altaffiliation[Present address:]{Fachbereich Physik, Unversit\"at
Konstanz, Universit\"atsstr.\ 10, 78457 Konstanz, Germany}

\affiliation{Experimentalphysik II, Universit\"at Augsburg,
Universit\"atsstr.\ 1, 86159 Augsburg, Germany}

\author{M. Dressel}
\affiliation{1.~Physikalisches Institut, Universit\"at
Stuttgart, Pfaffenwaldring 57, 70550 Stuttgart, Germany}
\author{M. Hanfland}
\affiliation{European Synchrotron Radiation Facility, BP 220, 38043
Grenoble, France}
\author{C. A. Kuntscher}\email{christine.kuntscher@physik.uni-augsburg.de}
\affiliation{Experimentalphysik II, Universit\"at Augsburg,
Universit\"atsstr.\ 1, 86159 Augsburg, Germany}

\date{\today}

\begin{abstract}
The infrared response of the organic conductor (TMTSF)$_2$\-PF$_6$ and the Mott insulator
(TMTTF)$_2$\-PF$_6$ are investigated
as a function of temperature and pressure and for the polarization
parallel and perpendicular to the molecular stacks.
By applying external pressure on (TMTTF)$_2$\-PF$_6$, the Mott
gap rapidly diminishes until the deconfinement transition occurs
when the gap energy is approximately twice the interchain transfer
integral. In its deconfined state (TMTTF)$_2$\-PF$_6$ exhibits a
crossover from a quasi-one-dimensional to a
higher-dimensional metal upon reducing the temperature.
For (TMTSF)$_2$PF$_6$ this dimensional cross\-over is observed
either with increase in external pressure or with decrease in
temperature. We quantitatively determine the dimensional crossover line in the
pressure-temperature diagram based on the degree of
coherence in the optical response perpendicular to the molecular stacks.
\end{abstract}


\maketitle

\section{Introduction}
The physics of one-dimensional (1D) conductors is substantially
different compared to two- or three-dimensional systems. Theory
shows that the effects of electronic interaction in an electron gas
have to be treated in a completely different way in one dimension
compared to three dimensions. The resulting non-Fermi-liquid metallic state is called the Luttinger
liquid (LL) state;\cite{Schulz91,Voit94,Giamarchi04}
it is characterized by a vanishing step in the occupation function leading to a 
power-law behavior in numerous physical quantities.
In the case of a half-filled conduction band Coulomb repulsion results in a 1D Mott insulating state.
These phenomena are realized in highly anisotropic systems
consisting of well-separated chains or stacks of structural entities.
In practice, however, a certain coupling between the 1D stacks is unavoidable, that 
can be described by a transverse hopping integral $t_{\bot}$.
Since this implies a possible charge transport normal to the chains, the system
becomes {\it quasi}-1D. If the interchain coupling is
strong enough, i.e., if $t_{\bot}$ exceeds a critical value
$t^*_{\bot}$, the Mott insulating state is
suppressed.\cite{Suzumura98,Tsuchiizu99a,Tsuchiizu99,Biermann01,Giamarchi04}
In other words: For $t_{\bot} < t^{*}_{\bot}$ the single-particle interchain
hopping is completely absent for $T=0$ and the system is a 1D Mott
insulator. In contrast, for $t_{\bot}>t^{*}_{\bot}$
interchain transport is present; the electrons are not
confined to the stacks any more and the system is in a quasi-1D metallic state.
The critical value $t^*_{\bot}$ for this  deconfinement transition is
given by the energy scale of the electronic correlations,
i.e., $t^*_{\bot}$ should be comparable to the value of the Mott gap
$\Delta_\rho\approx$$A\cdot$$t^*_{\perp}$ with the prefactor $A=1.8 - 2.3$.\cite{Suzumura98}

\begin{figure}[t]
  \centerline{
  \includegraphics[width=0.6\columnwidth]{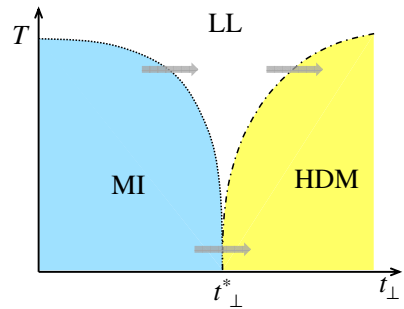}
  }
  \caption{(Color online) Schematic phase diagram of the deconfinement
  transition for a system of weakly coupled chains according to Refs.\
  \onlinecite{Biermann01} and \onlinecite{Giamarchi04}. The transition
  from a Mott insulator (MI) to a high-dimensional metallic (HDM) state
  occurs at $T=0$ when the transverse hopping integral
  $t_\perp$ reaches a critical value $t_\perp^*$. At high enough temperature,
  the increase in $t_\perp$ leads to a transition
  from a Mott insulator to a 1D Luttinger liquid (LL)
  state and, with further increase in $t_\perp$ to a dimensional crossover
  to a higher-dimensional metal (HDM).}\label{fig:diagram}
\end{figure}

Ideally, the deconfinement transition is a quantum phase transition that can
take place at zero temperature when the interchain coupling is varied.\cite{Giamarchi04}
The Mott insulator becomes a correlated metal (two or three
dimensional) due to the finite interchain coupling.
The 1D LL state is the consequence of the smearing of the Fermi-Dirac
distribution at high enough temperatures to mask the warping
of the open Fermi surface due to the interchain coupling $t_\perp$.
Hence, the deconfinement at temperatures
$T \gg t_\perp/k_B$ leads to a LL metallic state.
When the interchain coupling becomes stronger,
eventually a dimensional crossover from a LL to a higher-dimensional
metallic state is induced; it can also be reached when the temperature drops.
This dimensional crossover occurs when the warping of the Fermi surface
(proportional to $t_{\perp}$) exceeds the energy of thermal fluctuations,
i.e., $t_\perp > k_B T$.
The higher-dimensional metallic state can be described by
Fermi-liquid theory; its physical properties are distinctly different
compared to the LL state.\cite{Giamarchi-book}
All these phenomena are summarized in
the phase diagram sketched in Fig.~\ref{fig:diagram}.

Well-known models of quasi-1D systems are the Fabre salts (TMTTF)$_2X$ and the Bechgaard salts 
(TMTSF)$_2X$. \cite{Jerome82,Ishiguro98,Jerome04}
They consist of molecular stacks formed by TMTTF or TMTSF cations (which stands for 
tetra\-methyl\-tetra\-thia\-fulvalene and
tetra\-methyl\-tetra\-selena\-fulvalene, respectively).\cite{comment0}
Since half an electron is transferred from each organic molecule to the anions $X^-$, 
the band is three-quarter filled; the systems become
effectively half filled because TMTSF and TMTTF form  dimers. Charge transport 
occurs along the $a$ axis due to overlap of electronic orbitals
of sulfur or selenium atoms, respectively. The interstack separation 
is shortest in $b$ direction and determined by two facts: (i) the size of
the anions $X^-$ and (ii) the extension of the chalcogen orbitals. 
The (TMTTF)$_2X$ salts have weaker interstack coupling and therefore are 1D
Mott insulators. For the (TMTSF)$_2X$ salts, the interstack coupling is 
stronger resulting in a quasi-1D metallic state.

Applying external pressure to the crystals decreases
the intermolecular distances; the consequences are twofold:
(i) in stacking direction $a$ the bands widen slightly and (ii)
the interstack hopping integral $t_{\bot}$ increases exponentially.
Thus, it is expected that under pressure the electronic properties of 
(TMTTF)$_2X$ salts approach those of their TMTSF analogs. Over the years
extensive experimental investigations of Bechgaard-Fabre salts
made it possible to construct a generic temperature-pressure phase diagram \cite{Bourbonnais98,Wilhelm01}
containing a variety of ground states (Mott insulator,
spin-Peierls, Luttinger liquid, superconducting etc.).
According to their ambient-pressure properties,
all Bechgaard-Fabre salts with centrosymmetric anions (like AsF$_6$,
PF$_6$ or Br) can be located at certain position in the phase diagram.
The pressure offset between (TMTTF)$_2$PF$_6$ and (TMTSF)$_2$PF$_6$, for instance,
is about 3~GPa.\cite{Bourbonnais98} Thus, applying external pressure to the Mott
insulator \TF\ should induce a metallic state
similar to the one in \SF.

The first experimental evidence that relates the Mott gap $\Delta_{\rho}$ to the 
transverse hopping term $t^*_{\perp}$ at the deconfinement
transition was accomplished by comparing the Mott gap (obtained by ambient-pressure 
infrared spectroscopy) for various Bechgaard-Fabre salts
with anions of different size to the interstack hopping integrals estimated from 
tight-binding calculations.\cite{Vescoli98}  Vescoli {\it et al.} suggested that 
the deconfinement transition takes place around the region where 
$2 t_b \approx\Delta_{\rho}$. Obviously, due to the limited
number of compounds only a few points on the pressure axis can be simulated 
by chemical pressure. Moreover, the substitution of anions affects
not only the interstack coupling, but also the dimerization -- and thus 
the band structure -- along the stacks. A direct proof of the predicted
deconfinement transition and a quantitative determination of the dimensional 
crossover is highly desired.

External pressure is known to be a clean way to tune the relevant parameters 
of a system. Hence, infrared spectroscopy under external
pressure is the superior method to explore the deconfinement transition experimentally: 
Here, the changes in the electrodynamic response along
different directions can be monitored, while the interstack coupling is continuously 
raised. We recently applied this method to obtain a first
glance on the deconfinement transition in the Bechgaard-Fabre salts.\cite{Pashkin06,Pashkin06a} 
Since this study was restricted to room temperature and the
frequency range was limited, the size of the Mott gap as a 
function of pressure could only be inferred rather than directly obtained.

The one-dimensionality of the Bechgaard-Fabre salts has a crucial impact on 
their electronic properties in the deconfined state which is
described by the LL model. The dimensional crossover between quasi-1D and 
high-dimensional (Fermi liquid) electronic states shown in
Fig.~\ref{fig:diagram} is very interesting regarding theoretical\cite{Georges00,Biermann01,Giamarchi07}
and experimental\cite{Moser98,Auban04,Dressel05,Auban09} aspects.
Several power law exponents characterizing the energy and temperature
dependence of the conductivity change at the dimensional crossover.\cite{Dressel05,Giamarchi07} 
The best pronounced signature is expected for
the transverse resistivity in the direction of the weakest interchain coupling 
($c$ axis in case of the Bechgaard-Fabre salts). This transverse
charge transport can be described in terms of incoherent tunneling 
between the conducting chains. Since the density of states available for the
tunneling obey qualitatively different behavior in case of Luttinger and 
Fermi liquids the transverse resistivity is expected to show a
maximum at the dimensional crossover, which indeed has been 
observed in several experiments.\cite{Moser98,Auban04,Dressel05}
High-pressure measurements allowed Moser \textit{et al.}\cite{Moser98} 
to trace the dimensional crossover temperature $T^{*}$ as a function
of pressure $P$ for \SF. The observed $T^{*}(P)$ dependence is rather 
strong and non-linear pointing to a renormalization due to correlation
effects. The verification of the $T^{*}(P)$ dependence by an alternative 
method would be an important step in the characterization of the electronic
correlation effects in quasi-1D systems. As already pointed out by 
Jacobsen {\it et al.} \cite{Jacobsen81} optical data provide a convenient
possibility to determine the dimensionality of the electronic system, 
and thus to identify the dimensional crossover. Thus, probing the
interchain infrared response under external pressure offers a promising
alternative to the dc transport measurements.

In this paper we present the results of an extensive temperature-dependent 
infrared study of the organic salts \TF\ and \SF\ under external
pressure. These data combined with the pressure-dependent unit cell constants 
determined by x-ray diffraction measurements allowed us to directly obtain
the values of the Mott gap $\Delta_\rho$ and the interstack hopping integral 
$t_b$ as a function of pressure. Thus we were able to verify the
relation between $\Delta_\rho$ and $t_b$ at the point of the deconfinement 
transition in \TF. Furthermore, we estimate the line of the
dimensional crossover in the temperature-pressure diagram of \TF\ and \SF. 
Based on the obtained results, we quantitatively suggest a unified
phase diagram for the studied Bechgaard-Fabre salts.

The manuscript is organized as follows. Section~\ref{Experiment} 
gives a brief description of the experimental techniques. The x-ray
diffraction data are shortly presented in Section~\ref{XRD}. 
Section~\ref{Infrared} contains the infrared spectroscopy results. 
The pressure-dependent values of the Mott gap and the transverse 
hopping integral are given in Section~\ref{deconfinement} and 
discussed in terms of the pressure-induced deconfinement transition. 
Section~\ref{crossover} discusses the infrared response along the $b'$ 
direction in terms of the interstack transport coherence. 
The dependence of the so-called coherence parameter on pressure and temperature 
is presented and the lines of the dimensional crossover 
are determined. Finally, Section~\ref{Summary} summarizes the results 
within a unified temperature-pressure phase diagram.

\section{Experiment}\label{Experiment}

Single crystals of \TF\ and \SF\ were grown by a standard
electrochemical growth procedure. The measured samples had as-grown specular surfaces
in $ab$ plane and a thickness of about 50 $\mu$m.
The room-temperature polarized infrared reflectivity was measured
in the range 200 - 8000 \icm\ using a Bruker IRscope II microscope
attached to a Bruker IFS66v/S Fourier transform
spectrometer. The measurements in the far-infrared range (200 -
700 \icm) were performed at the infrared beamline of the
synchrotron radiation source ANKA. A diamond anvil cell (DAC)\cite{DAC}
equipped with type IIA diamonds suitable for infrared measurements
was used to generate pressures up to 6~GPa. Finely ground CsI
powder was chosen as quasi-hydrostatic pressure transmitting
medium. 

The mid-infrared reflectivity (580 - 8000 \icm) at low temperature
and high pressure was measured using a home-built infrared
microscope coupled to the spectrometer and maintained
at the same vacuum conditions, in order to avoid
absorption lines of H$_2$O and CO$_2$ molecules. The infrared
radiation was focused on the sample by all-reflecting home-made
Schwarzschild objectives with a large working distance of about
55~mm and 14x magnification. The DAC was mounted on the
cold-finger of a continuous flow helium cryostat (Cryo\-Vac
KONTI) with two metallic rods, which allow
mechanical access to the DAC lever arm mechanism. Thus, the
pressure in the DAC could be changed {\it in situ} at arbitrary
temperature.

In all high-pressure measurements the pressure in the DAC was determined
{\it in situ} by the ruby luminescence method.\cite{Mao86}
More details about the geometry of the reflectivity measurements and the Kramers-Kronig
(KK) analysis of the data can be found in our earlier
publications.\cite{Pashkin06,Kuntscher06}

In addition we carried out room-temperature x-ray
diffraction (XRD) experiments at beamline ID09A of the European Synchrotron Radiation
Facility at Grenoble. Liquid helium served
as pressure transmitting medium. The DAC rotation angle
varied from $-30^{\circ}$ to $+30^{\circ}$ for (TMTTF)$_2$PF$_6$
and from $-20^{\circ}$ to $+20^{\circ}$ for (TMTSF)$_2$PF$_6$ with
$2^{\circ}$ step. The diffraction patterns of the single crystals 
have been analyzed using the XDS package.\cite{Kabsch93}
The details are given in Ref.~\onlinecite{Pashkin09}.

\begin{figure}[t]   
  \centerline{
  \includegraphics[width=0.7\columnwidth]{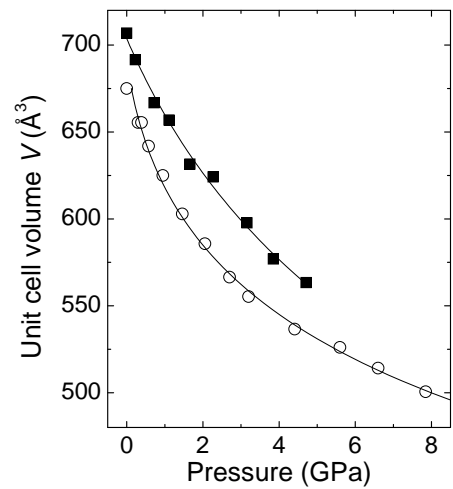}
  }\caption{Pressure-dependent unit-cell volume of \TF\ (open circles) and \SF\ (full rectangles) 
  at room temperature.
  The lines represent the fits according to the Birch equation of state, Eq.\ (\ref{Birch equation}).}
  \label{fig:unitcell}
\end{figure} 

\section{Results}

\subsection{X-ray diffraction}\label{XRD}

The single-crystal XRD data show that the space-group symmetry
$P\bar{1}$ of both studied compounds\cite{Delhaes79,Gallois86} remains 
intact up to pressures of 6-8 GPa.\cite{Pashkin09} The dependence of
unit-cell volume of (TMTTF)$_2$PF$_6$ and (TMTSF)$_2$PF$_6$ is shown
in Fig.~\ref{fig:unitcell}. The change of the unit-cell volume can
be fitted with the Birch equation of state:\cite{Birch78} 
\begin{equation} \label{Birch equation}
P(V)=\frac{3}{2}B_0(x^7-x^5) \cdot
\left[1+\frac{3}{4}(B'_0-4)(x^2-1)\right] \quad ,
\end{equation}
with $x=(V/V_0)^{1/3}$
and $V_0$ denotes the unit cell volume at ambient pressure. The
values of the bulk modulus $B_0$ obtained from the fits are $7.27
\pm 0.64$~GPa and $12.71\pm 0.97$~GPa for \TF\ and \SF,
respectively. The corresponding pressure derivatives of the bulk
modulus $B'_0$ are $9.98 \pm 1.15$ and $4.23 \pm 0.75$,
respectively. Our findings agree with previous structural studies of \SF\ under moderate
pressure.\cite{Gallois86,Gallois87}

The Bechgaard-Fabre salts exhibit a strong thermal expansion\cite{Gallois86} 
that has to be taken into account, for instance, when dc transport
experiments are compared with theoretical predictions of some power-law 
behavior on $\rho(T)$. \cite{Korin-Hamzic03,Dressel05} We acknowledge
this dependence also in our present analysis of the optical conductivity 
(see Section~\ref{Infrared}). The underlying idea is that the reduction
of the unit-cell volume is achieved either by applying pressure or by 
reducing the temperature. Utilizing the temperature dependence of the
unit-cell parameters of \SF\ measured at ambient pressure\cite{Gallois86} 
we obtain a relation between temperature and the pressure of
$2.3\times 10^{-3}$~GPa/K, assuming linearity. This means, for examples, 
that cooling down by 100~K at ambient pressure is equivalent to
applying 0.23~GPa pressure. By assuming that this relation is also 
valid at high pressures and the sulfur analog, we were able to estimate
the unit-cell volume at any point of the temperature-pressure phase 
diagram for the analysis of our infrared spectroscopic results.

\begin{figure}
  \centerline{
  \includegraphics[angle=0,width=0.85\columnwidth]{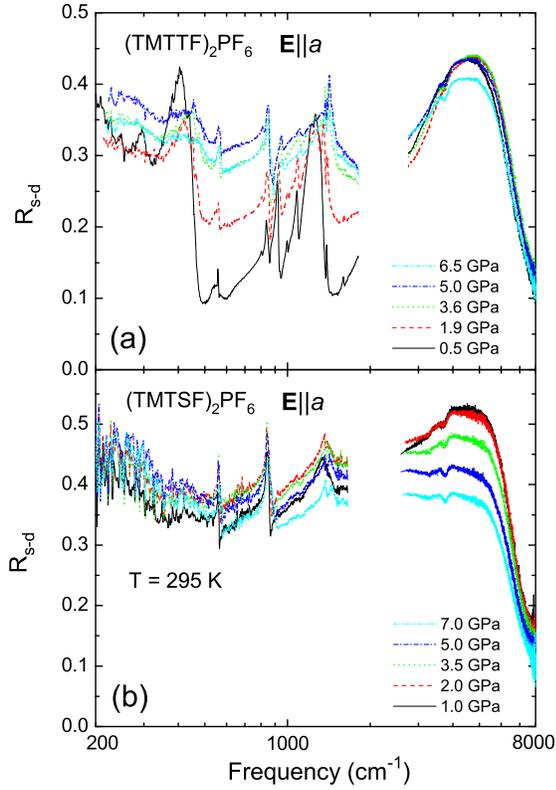}
  }\caption{(Color online) Pressure-dependent reflectivity
  spectra $R_{s-d}$ of \TF\ and \SF\ measured at room temperature 
  for the polarization \textbf{E}$\parallel a$.}
  \label{fig:drefl-a}
\end{figure}
\begin{figure}
  \centerline{
  \includegraphics[width=0.9\columnwidth]{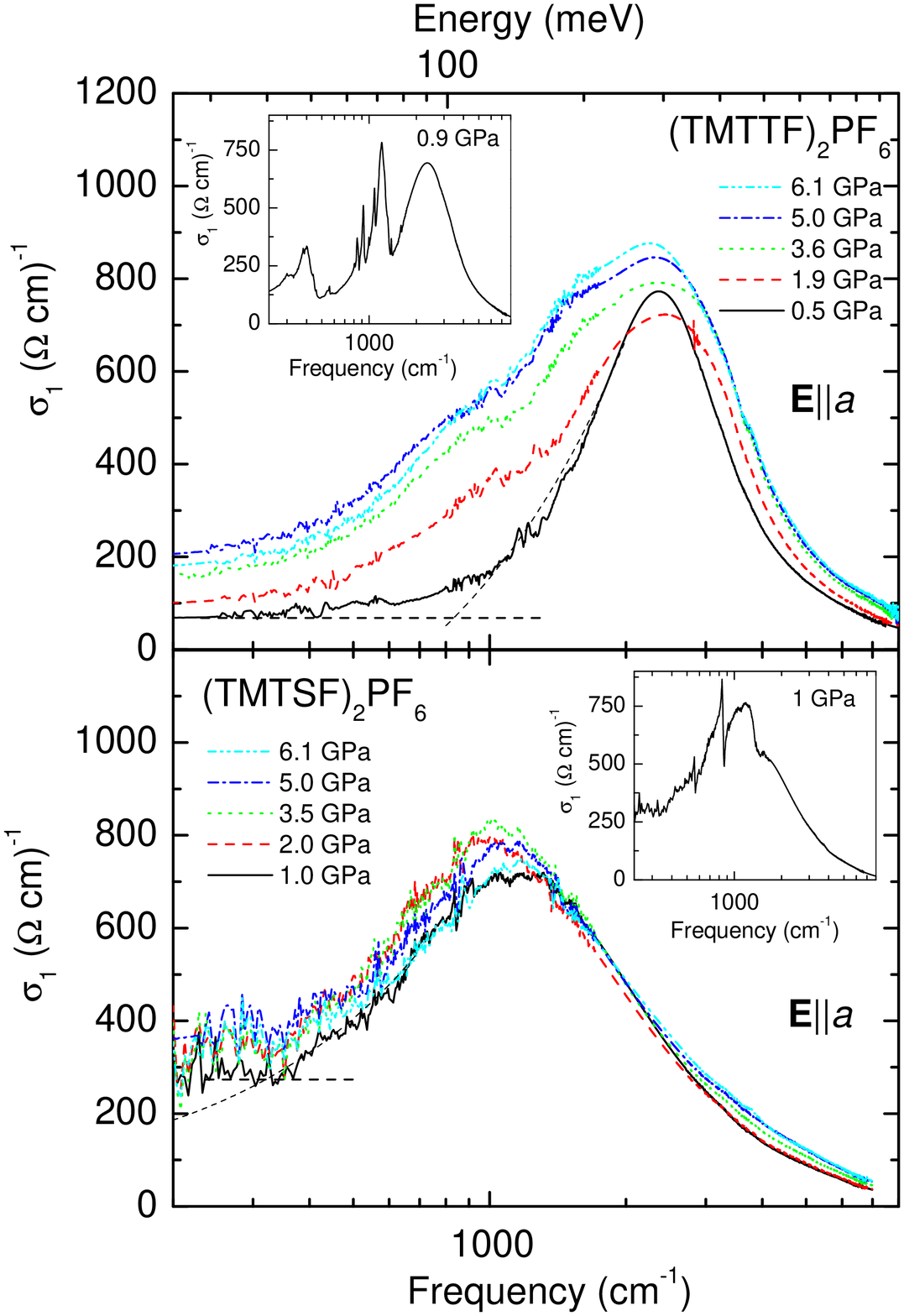}
  }\caption{(Color online) Real part of the optical conductivity $\sigma_1$ 
  of \TF\ and \SF\ obtained from the room-temperature reflectivity for the
  polarization ${\bf  E} \parallel a$.  For clarity reason the
  phonon contributions are subtracted;
  examples of the complete conductivity spectra of \TF\ and \SF\ at low pressure 
  are plotted in the insets. The dashed lines illustrate the method used to determine
  the Mott gap.
  }\label{fig:sigma-a}
\end{figure}

\subsection{Infrared optical response}\label{Infrared}
\subsubsection{Parallel to the stacks}
\label{sec:Intra-stack}

Fig.~\ref{fig:drefl-a} displays the room-temperature reflectivity spectra
measured at the sample-diamond interface, $R_{s-d}$, for light polarized along the stack (\textbf{E}$\|a$).
The data extend down to 200~\icm\ by combining spectra measured with a standard globar 
source with those using the synchrotron radiation. At around
2000~\icm\ the reflectivity is strongly influenced by multiple phonon
absorption in the diamond anvil that causes artifacts in the spectra.
Therefore, this frequency region is cut out and interpolated for the analysis.
For \TF\ the reflectivity spectra [Fig.~\ref{fig:drefl-a}(a)]
consist of a strong mid-infrared (mid-IR)
band and a number of sharp vibrational modes below 2000~\icm, which correspond to
electron-molecular-vibration (emv) coupled totally symmetric vibrations of
the TMTTF molecules and infrared-active modes of the PF$_6$
anion.\cite{Jacobsen83} The strong pressure dependence of $R_{s-d}$
agrees with our previous findings on the sister compound \asf.\cite{Pashkin06,Pashkin06a}
Most important, the low-frequency reflectivity (below 1700~\icm)
increases with pressure when the Mott gap vanishes; at the same time the
vibrational modes shift to higher frequencies and broaden
considerably.

In the case of \SF, the reflectivity spectra show similar features as observed 
for \TF\ at pressures above 3~GPa [Fig.~\ref{fig:drefl-a}(b)
compared to Fig.~\ref{fig:drefl-a}(a)]. Interestingly, the emv modes are less 
pronounced compared to TMTTF because the dimerization of the stacks in \SF\ 
is much weaker.\cite{Jacobsen83,Pedron94} The pressure-induced changes 
of the \SF\ reflectivity are rather small and mainly consist in a decrease
of the mid-IR feature.

The corresponding room-temperature conductivity spectra are
obtained from the ${\bf E}\parallel a$ reflectivity by means of
Kramers-Kronig analysis. The real part of the optical conductivity
$\sigma_1$ of \TF\ and \SF\ at low pressure is shown in the insets
of Fig.~\ref{fig:sigma-a}. For both materials the most pronounced feature
is an absorption band in the mid-IR due to excitations of charge carriers
from the lower to the upper Hubbard band. In order to disentangle the electronic
and vibrational contributions to the optical conductivity, we fit
the spectra with the Drude-Lorentz model, taking into account the
Fano lineshapes\cite{Fano61} of the emv-coupled modes. Then, the
phonon contributions are subtracted and the resulting conductivity
spectra are plotted in Fig.~\ref{fig:sigma-a} for selected 
pressures.\cite{comment3}
With increasing pressure the ${\bf E} \parallel a$ optical conductivity of \TF\ 
undergoes considerable changes, with a large increase in spectral
weight in the frequency range $<$8\,000~\icm\ due to the enlarged
bandwidth. The enhancement is linear with pressure but saturates
above 5~GPa (not shown). In addition, the spectral weight, as determined
by the center of gravity of $\sigma_1(\omega)$ in the region plotted in 
Fig.~\ref{fig:sigma-a}(a), shifts to lower energies.
Accordingly, the value of the Mott gap decreases. The size of
the Mott gap $\Delta_\rho$ is estimated by the intersection  between the 
linear extrapolation of the low-frequency edge of the Mott-Hubbard band 
and the constant background level of low-frequency conductivity, as 
illustrated by the dashed lines in Fig.~\ref{fig:sigma-a}. 
In contrast, the optical conductivity of \SF\ exhibits
only minor changes with pressure; the Mott-Hubbard band and the
free-carrier motion along the molecular stacks are almost pressure
independent.
The pressure dependence of the Mott gap $\Delta_{\rho}$  for \TF\ and \SF\
is depicted in Fig.~\ref{fig:deconf} and will be discussed in detail in 
Section \ref{deconfinement}.

\begin{figure}[t]
  \centerline{
  \includegraphics[angle=0,width=0.85\columnwidth]{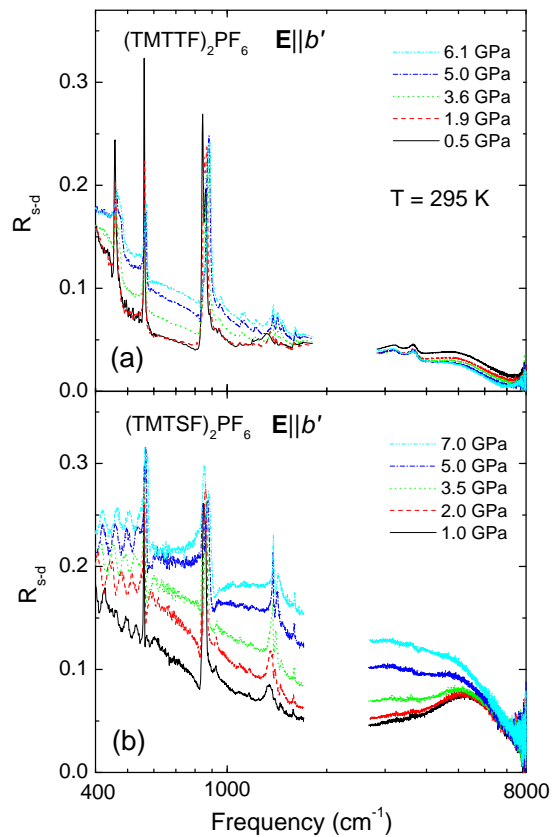}
  }\caption{(Color online) Pressure-dependent reflectivity
  spectra $R_{s-d}$ of \TF\ and \SF\ measured at room temperature 
  for the polarization \textbf{E}$\parallel b'$.
}\label{fig:reflectivity-b}
\end{figure}

\begin{figure}
  \centerline{
  \includegraphics[angle=0,width=0.8\columnwidth]{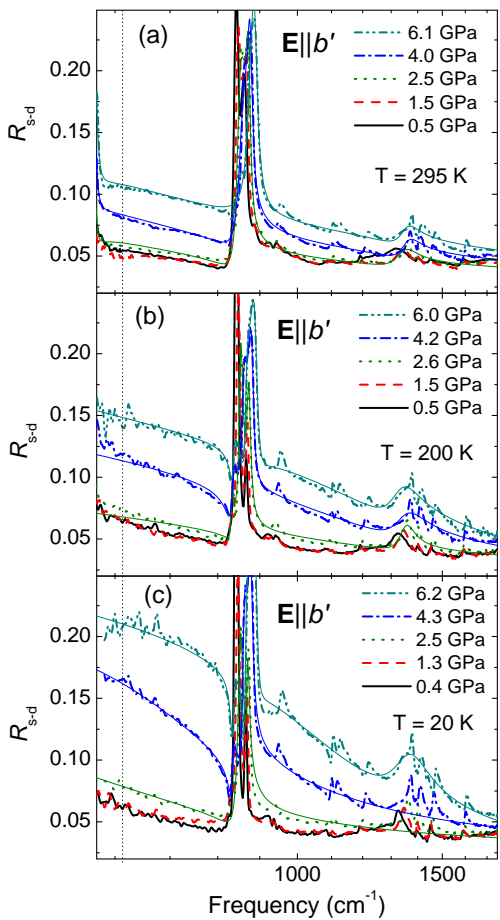}
  }\caption{(Color online) Pressure-dependent reflectivity spectra $R_{s-d}$ 
  of \TF\ for the electric field polarized
normal to the stacks, \textbf{E}$\parallel b'$, (a) at room
temperature, (b) at 200 K and (c) at 20 K.
The vertical dotted line marks the frequency of 610~\icm\ at which the reflectivity
change $\delta R_{s-d}$ is estimated (see Fig.~\protect\ref{fig:levels}).
The solid lines are fits with a Drude term combined with Fano oscillators.}
\label{fig:dreflb}
\end{figure}

\subsubsection{Perpendicular to the stacks}
\label{sec:Inter-stack}

In Fig.~\ref{fig:reflectivity-b} we show the reflectivity spectra of \TF\ 
and \SF\ measured for ${\bf E}\parallel b'$ at selected pressures. In general the 
absolute value is significantly lower compared to the corresponding data along the 
stacks, and $R_{s-d}$ barely exceeds 0.25 in the measured spectral range. 
At the lowest applied pressure the
reflectivity spectrum of \TF\ is almost flat in the mid-infrared range
with an increase in the far-infrared [Fig.~\ref{fig:reflectivity-b}(a)].
The response does not change with pressure up to 2~GPa. In this
range, \TF\ remains in the confined state and the coherent
interstack  transport is suppressed.\cite{comment4}
Only above 2~GPa the overall reflectivity increases
and the spectrum develops a shape typical for a conductor; i.e.\ the
reflectivity increases towards lower frequencies. This indicates
the deconfinement of the interstack charge transport: Above the
critical pressure a coherent Drude response sets in for ${\bf
E}\parallel b'$. \SF\ [Fig.~\ref{fig:reflectivity-b}(b)] is in the deconfined 
state already at ambient pressure;  thus the reflectivity continuously 
increases as pressure is applied. In contrast to the \TF\ salt, no threshold 
at a critical pressure is observed in case of \SF.

The Drude response in the deconfined state of \TF\ (i.e., above 2~GPa) changes its character
with temperature. As seen in Fig.~\ref{fig:dreflb}(a) for $T=295$~K
the low-frequency reflectivity increases only slightly with decreasing frequency,
which is typical for an overdamped
Drude response when the relaxation rate exceeds the plasma frequency.
In contrast, the Drude behavior gets more pronounced in the low-temperature
reflectivity spectra [Fig.~\ref{fig:dreflb}(b) and(c)];
the low-frequency reflectivity strongly increases with decreasing frequency 
and a well-defined plasma edge develops.

\begin{figure}
  \centerline{
  \includegraphics[angle=270,width=0.8\columnwidth]{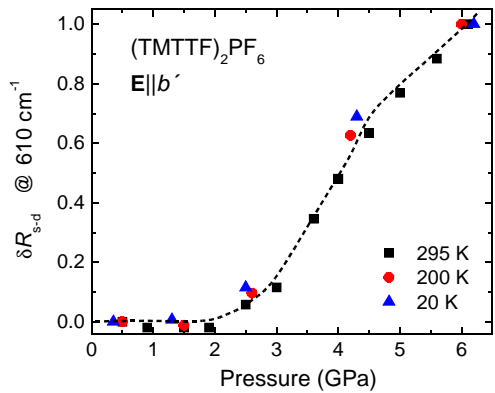}
  }\caption{(Color online) Normalized pressure-induced change of the reflectivity 
  level of \TF\ for ${\bf E}\parallel b^{\prime}$ at 610
  \icm\ at different temperatures as indicated. The dashed line is a guide to the eye.} 
  \label{fig:levels}
\end{figure}

Fig.~\ref{fig:levels} displays the normalized changes of the reflectivity values 
$R_{s-d}$ observed at 610~\icm\ as a function of pressure for three different
temperatures. The reflectivity change is defined as $\delta
R_{s-d}=(R_{s-d}-R_{s-d}^l)/(R_{s-d}^h-R_{s-d}^l)$, where
$R_{s-d}^l$ and $R_{s-d}^h$ are the sample-diamond reflectivity at the
lowest and highest applied pressures. Clearly, 
$\delta R_{s-d}$ starts to increase only above a certain threshold:
The deconfinement transition in \TF\ can unambiguously be identified
at about 2~GPa. The coherent response leading to this 
Drude behavior increases as the temperature is reduced, the critical pressure of the
deconfinement transition remains basically temperature-independent.

\begin{figure}[t]
  \centerline{
  \includegraphics[angle=0,width=0.9\columnwidth]{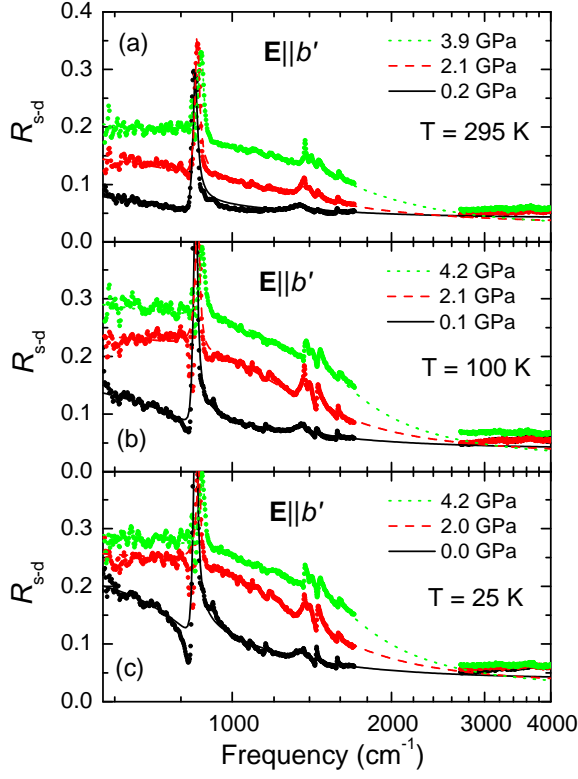}
  }\caption{(Color online) Pressure-dependent reflectivity spectra $R_{s-d}$ 
  of \SF\ for \textbf{E}$\parallel b'$ (a) at
room temperature, (b) at 100 K and (c) at 25 K. The solid lines are fits with a Drude
term combined with Fano oscillators.}\label{fig:sf-b}
\end{figure}

For the Bechgaard salt \SF\ the interstack transport changes continuously with
pressure or temperature as seen from reflectivity spectra in Fig.~\ref{fig:sf-b}. 
The Drude-like optical response at the
lowest applied pressure becomes more pronounced on cooling down
indicating the onset of coherent electronic transport between the
stacks. This observation agrees with earlier
ambient-pressure infrared studies\cite{Jacobsen81,Degiorgi96,Vescoli99a}
and it is discussed quantitatively in Section~\ref{crossover}.
At pressures above 2~GPa and for temperatures below 100~K a plasma edge
is clearly observed in the mid-infrared range [Fig.~\ref{fig:sf-b}(c)].

\begin{figure}
  \centerline{
  \includegraphics[width=0.9\columnwidth]{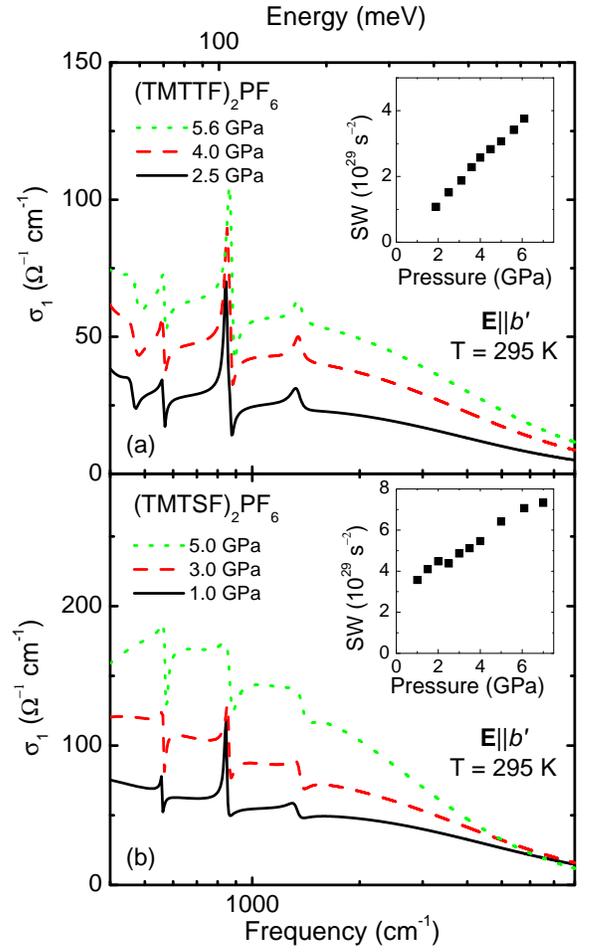}
  }\caption{(Color online) Real part of the optical conductivity $\sigma_1$ 
  of \TF\ (a) and \SF\ (b) obtained
  from the fits of room-temperature reflectivity for the
  polarization ${\bf  E} \parallel b'$. Insets show the respective spectral weight 
  obtained according to Eq.\ \ref{SW} as a function of pressure.}
   \label{fig:sigma-b}
\end{figure}

For a quantitative analysis of the transverse response, we fit the
reflectivity spectra of of \TF\ and \SF\ with the Drude model
\begin{equation}
\sigma(\omega) = \frac{\epsilon_0 \omega_p^2}{\Gamma-i\omega}
\label{eq:Drude}
\end{equation}
for the coherent transport and several Fano oscillators \cite{Fano61} 
describing intra-molecular vibration modes. In Eq.~(\ref{eq:Drude}) the
plasma frequency is denoted by $\omega_p$ and the scattering rate by $\Gamma$. 
The fitting curves for the corresponding experimental ${\bf
E}\parallel b'$ reflectivity spectra are displayed in Figs.~\ref{fig:dreflb} 
and \ref{fig:sf-b}. From the fits, we obtained the
real part of the optical conductivity for both salts, which is 
presented in Fig.~\ref{fig:sigma-b} 
for several selected pressures. The room temperature spectra of \TF\
are fitted only in the deconfined phase (i.e., above 2~GPa), since the Drude 
response is supposed to be absent when the electrons are confined 
to the stacks. It is clear from Fig.~\ref{fig:sigma-b} that for both 
salts the interstack conductivity strongly increases with pressure. 

We also consider the pressure-dependent spectral weight, calculated according to 
\begin{equation} \label{SW}
SW=\int_0^{\omega_c} \sigma_1(\omega){\rm d}\omega
\end{equation}
for ${\bf E}\parallel b'$ as a function of pressure, where the
cut-off is chosen as $\omega_c=8000~{\rm cm}^{-1}$. As expected, the 
spectral weight of \TF\ is significantly smaller compared to \SF, however, it shows a
stronger pressure dependence (see insets of Fig.\ \ref{fig:sigma-b}). 
Already at ambient pressure the
electrons are not confined in the stacks of \SF;
consequently, the spectral weight is relatively high for \textbf{E}$\parallel b'$. 
Furthermore, its pressure dependence is weaker and it saturates at high pressures. 
The values of the spectral weight allows us to estimate the transverse hopping
integral $t_b$, as discussed in Section~\ref{deconfinement}.
Furthermore, the parameters of the Drude response for ${\bf E}\parallel b'$ 
show the signatures of the expected temperature- and pressure-induced 
dimensional crossover, which will be considered in Section~\ref{crossover}.

\section{Discussion}

\subsection{Deconfinement transition}\label{deconfinement}

The interstack infrared reflectivity spectra displayed in
Fig.\ \ref{fig:dreflb} provide clear indications of the
pressure-induced deconfinement transition in \TF\ at 2~GPa, as
summarized in Fig.~\ref{fig:levels}. We now compare our experimental
findings with the theoretical predictions for the deconfinement transition, 
in particular the relation between the Mott gap energy $\Delta_\rho$ and the
interstack transfer integral $t_b$.

\begin{figure}[t]
  \centerline{
  \includegraphics[angle=0,width=1\columnwidth]{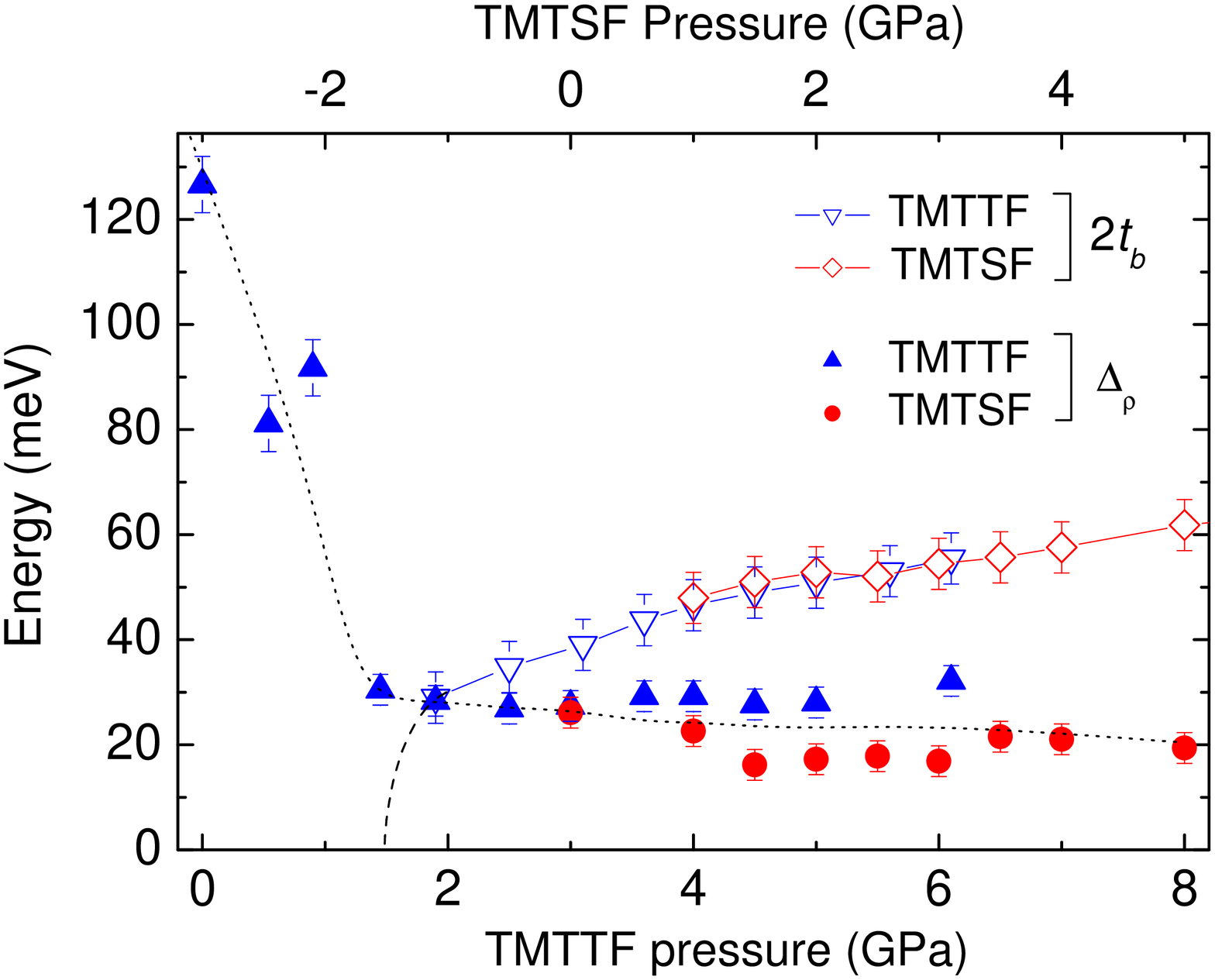}
  }\caption{(Color online) Pressure dependence of the Mott gap
  energy $\Delta_\rho$ and the
  transverse hopping integral $2t_b$ for \TF\ and \SF\ salts as obtained from the room-temperature data.
  The upper pressures scale corresponds to \SF\ and is shifted by 3~GPa with respect 
  to the lower pressure scale for the \TF\ salt. The dashed line indicates how $t_b$ vanishes  
  below the deconfinement transition.
  }\label{fig:deconf}
\end{figure}

First let us consider the in-stack optical response: The evolution of $\Delta_\rho$ 
as a function of pressure, determined from the ${\bf E}\parallel a$
conductivity spectra as described in Section \ref{sec:Intra-stack}, 
is displayed in Fig.~\ref{fig:deconf} for both salts under investigation. 
The dramatic decrease in $\Delta_\rho$ in \TF\ on
approaching the deconfinement transition at 2~GPa is in excellent agreement 
with theoretical studies \cite{Berthod06} and supported by earlier
experiments.\cite{Auban04} It is important to note that the gap does not 
vanish completely above the transition; this was already inferred by
earlier investigations on \SF.\cite{Dressel96} The theory of the Mott 
transition in quasi-1D systems predicts that the gap may persist in the
metallic phase.\cite{Giamarchi97,Schwartz98} By comparing both compounds 
under investigation, the values of the Mott gap for \SF\ match well
with the data for \TF, if we take into account a pressure offset of 3.0~GPa 
between the two salts; this value is in accordance with the generic
temperature-pressure phase diagram of the Bechgaard-Fabre salts.\cite{Bourbonnais98}

Next we discuss the optical response perpendicular to the stacks: 
The pressure-dependent values of the interstack transfer
integral $t_b$ are obtained by the quantitative analysis of the
transverse optical response. To this end we calculate the values of
$t_b$ using the plasma frequencies $\omega_p$
obtained from the Drude fits.
At ambient conditions, both salts are quasi-1D compounds with an open Fermi surface,
and the transverse hopping integral $t_b$ is given by\cite{Kwak82}
\begin{equation}
  t_b^2 = \frac{\pi \epsilon_0 \hbar^2 V_c t_a \omega_p^2}{4 e^2 b^2} \quad,
  \label{eq:Kwak}
\end{equation}
where $V_c$ denotes the unit cell volume, $b$ the separation of the stacks, 
and $t_a$ the transfer integral along the stacks. The geometrical
parameters $V_c$ (Fig.~\ref{fig:unitcell}) and $b$ were obtained by 
pressure-dependent x-ray diffraction measurements.\cite{Pashkin09} The value of $t_a$
was calculated from the spectral weight of the in-stack optical 
conductivity using a simple tight-binding model.\cite{Pashkin06}

The calculated transverse hopping integral $t_b$ is depicted in Fig.~\ref{fig:deconf}
as a function of pressure. The values of $t_b$ for \TF\ and \SF\ perfectly agree with each other, 
taking into account a pressure offset of 3~GPa between these compounds.\cite{comment1}
Above the deconfinement transition (i.e., $P>2$~GPa) and up to 4~GPa the increase
in $t_b$ for \TF\ is almost linear with a slope of 5~meV/GPa, in accord with
our earlier results for \asf.\cite{Pashkin06,Pashkin06a}
Above 4~GPa the pressure-induced increase in $t_b$ becomes weaker ($\sim$2~meV/GPa) 
for \TF\ and comparable to the low-pressure behavior of \SF.

Finally, the quantitative criterion $\Delta_{\rho}\approx 2\,t^*_b$ for the deconfinement
transition \cite{Suzumura98} can be verified by comparing the pressure dependence
of the charge gap $\Delta_\rho$ with that of 2$t_b$ (see Fig.~\ref{fig:deconf}). 
The onset of the coherent electronic transport normal to the stacks occurs at
around 2~GPa, where $\Delta_\rho \approx 2 t_b$. This agrees very
well with our conclusions drawn from the pressure-dependent reflectivity results shown 
in Fig.~\ref{fig:dreflb}. Thus, the results of our pressure-dependent
infrared spectroscopic study exhibit an overall consistency and support the theoretical 
picture \cite{Suzumura98,Biermann01,Giamarchi04} for the deconfinement
transition in quasi-1D compounds.

It is interesting to compare our results for the deconfinement transition induced by 
{\it external} pressure with the earlier findings obtained for anion substitution, 
i.e., the {\it chemical} pressure effect:\cite{Vescoli98} Based on ambient-pressure infrared
measurements, the Mott gap was deduced for several Bechgaard-Fabre salts and compared to 
the values of the interstack hopping integrals estimated from tight-binding calculations. 
The results strongly support our conclusion and the theoretical predictions that 
$2 t_b \approx\Delta_{\rho}$ at the deconfinement transition.

\subsection{Dimensional crossover}\label{crossover}

According to Section~\ref{sec:Inter-stack}, for both compounds in the deconfined state 
one observed an enhancement of the Drude-like response along the $b'$ direction 
when pressure is applied. Concomitant
with the compression of the crystal lattice the orbital overlap between neighboring molecular
stacks increases, and hence the Drude response along the $b'$ direction becomes stronger. 
On the other hand, upon lowering the temperature the scattering rate decreases, with a 
concomitant narrowing of the Drude term. Thus, at high pressure and low temperature 
the interstack transport is expected to be coherent and the dimensionality of the 
system should be enhanced (see Fig.~\ref{fig:diagram}). This dimensional crossover has drawn
considerable interest in the theory of correlated electron systems
of reduced dimensions.\cite{Biermann02}

\begin{figure}
  \centerline{
  \includegraphics[width=0.9\columnwidth]{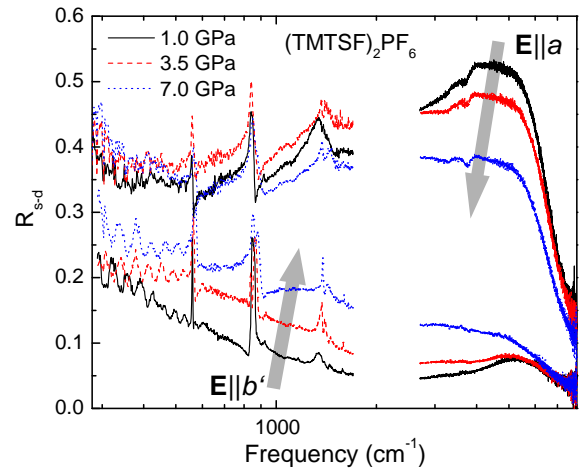}
  }\caption{(Color online) Comparison of the room-temperature reflectivity spectra $R_{s-d}$ 
  of \SF\ along and perpendicular to the stacks as pressure increases. The arrows indicate the 
  major changes in the spectra with increasing pressure.}
  \label{fig:SF-crossover}
\end{figure}

The trend towards a higher dimensionality with increasing pressure is illustrated by a comparison
of the room-temperature infrared reflectivity spectra of \SF\ for the polarization along and 
perpendicular to the stacking direction (see Fig.~\ref{fig:SF-crossover}). Obviously, 
the anisotropy decreases  with pressure: the Drude response starts to 
dominate the reflectivity for both polarizations and simultaneously 
the Mott-Hubbard band becomes less pronounced at high pressures. At the highest pressure (7~GPa) the 
reflectivity spectra along and perpendicular to the stacks are rather similar regarding the
principle shape, especially their low-frequency part (below 1000~\icm). Thus, \SF\ can readily 
be considered as a two-dimensional system at pressures around 7~GPa.

\begin{figure}
   \centerline{\includegraphics[width=0.8\columnwidth]{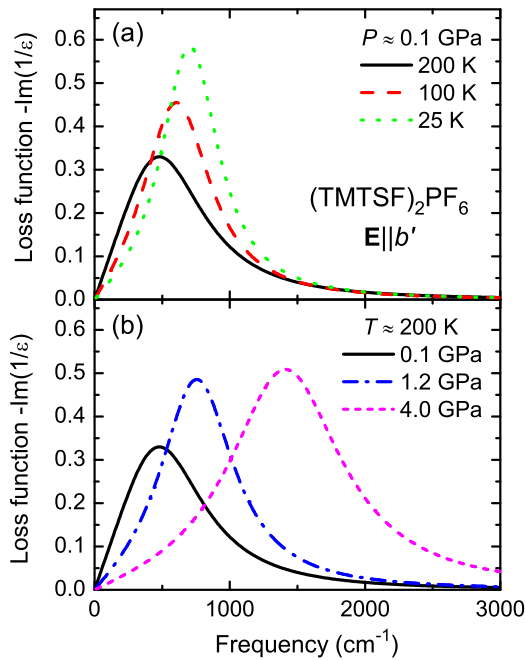}}
  \caption{(Color online) Loss function of the Drude response perpendicular to the conducting
  stacks (${\bf E} \parallel b'$) in \SF:
   (a) at 0.1~GPa and three selected temperatures; (b) at 200 K and three selected pressures.}
  \label{fig:loss}
\end{figure}

For a quantitative analysis of the dimensional crossover we now consider the degree of 
coherence of the charge transport. The hallmark of coherence is a
Drude response with a plasma frequency exceeding the damping constant.\cite{comment2} 
One can define a coherence parameter of the Drude response
\begin{equation}
\kappa = \frac{\omega_p }{ 2\Gamma} \quad ,
\label{kappa}
\end{equation}
where $\omega_p$ and $\Gamma$ are the plasma frequency and the scattering
rate of the Drude term defined in Eq.~(\ref{eq:Drude}).
Coherent transport corresponds to $\kappa > \kappa^*$, while incoherent transport
occurs for $\kappa < \kappa^*$, where $\kappa^*$ is a critical
value of the dimensional crossover which should be close to unity.
The factor 2 in Eq.~(\ref{kappa}) stems from the analogy between
the loss function of the Drude response and the
response of a simple harmonic oscillator which is considered
overdamped when its eigenfrequency is smaller than twice the damping
constant.\cite{Lucovsky76}

The above approach can be illustrated by the loss function
\begin{equation}
-{\rm Im}\left\{\frac{1}{\epsilon(\omega)}\right\} = \frac{\omega_p^2\omega\Gamma}{(\omega^2-\omega_p^2)^2+\omega^2\Gamma^2}
\end{equation}
presented in Fig.~\ref{fig:loss} for various temperatures and pressures.
In case of a Drude response the loss function has a Lorentzian line shape where the peak 
position corresponds to the plasma frequency $\omega_p$ and the width of the peak
corresponds to the electron scattering rate $\Gamma$.\cite{Dressel-book} Thus, the sharpness 
of this so-called plasmon peak in the loss function characterizes the degree of coherence of the
charge transport. Fig.~\ref{fig:loss}(a) demonstrates the evolution of the
plasmon peak for ${\bf E} \parallel b'$ in \SF\ at the lowest
applied pressure at selected temperatures. For 200~K the peak is
clearly overdamped. Upon cooling its width decreases, resulting in a well-developed
plasmon at 25~K. The plasmon peak exhibits a small blueshift with temperature
decrease. Its evolution with pressure is significantly different: For a fixed
temperature the plasmon peak strongly shifts to higher frequencies, while the peak width
only slightly increases [see Fig.\ \ref{fig:loss}(b)]. Thus, also the pressure application
results in an underdamped plasmon peak. In conclusion, there are two ways to tune the
dimensionality of the system and hence induce a dimensional crossover to an HDM state -
either by decreasing the temperature or by increasing the pressure. 

\begin{figure}[t]
  \centerline{
  \includegraphics[width=1\columnwidth]{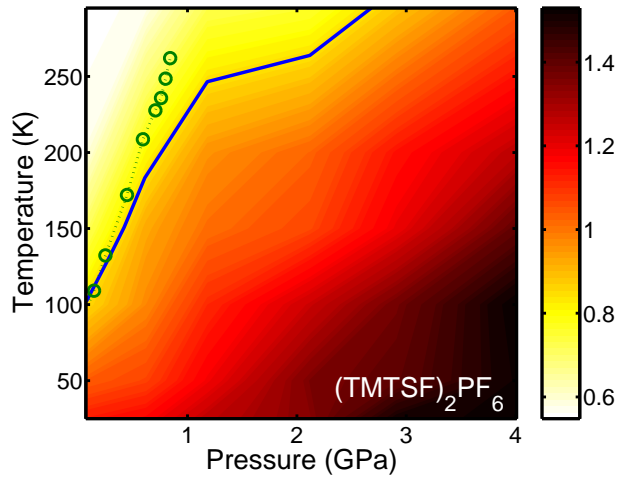}
  }\caption{(Color online) The coherence parameter of the interstack charge transport $\kappa_b$
  as a function of temperature and pressure
  for \SF. The solid blue line corresponds to $\kappa_b = 0.85$.
  The circles illustrate the line of the dimensional crossover
  determined from transport measurements by Moser {\em et al.}\cite{Moser98} The dotted line is a guide to the eye.
  }\label{fig:coherenceSF}
\end{figure}

The evolution of the dimensionality of \SF\ is summarized in
Fig.~\ref{fig:coherenceSF}, where we plot the coherence parameter $\kappa_b$ for \SF\ 
as a function of temperature and pressure, determined from the Drude fits of the interstack 
reflectivity spectra. The large number of experimental data points in the
diagram (six temperatures for each of the six pressure values) provides a 
rather detailed information on how the coherent transport develops.
The highest degree of coherence is achieved for high pressures and low
temperatures (lower right corner of the diagram). It gradually
decreases as the pressure is released and the temperature is simultaneously raised
towards the upper left corner of the diagram. The lines of constant coherence level 
are almost linear with pressure for $\kappa_b > 1$, however, for
$\kappa_b < 1$ they become sublinear. In order to define the
critical $\kappa^*$ values that characterize the dimensional
crossover, we utilize the experimental results of Moser {\it et
al.}\cite{Moser98}, where the dimensional crossover temperature in \SF\ at 
ambient pressure is proposed to be located at $T^* \approx 100$~K,
in accordance with other dc and microwave measurements of the $c^*$
axis resistivity.\cite{Cooper86,Dressel05} Using this as a reference
point, we obtain the critical coherence parameter $\kappa_b \simeq
0.85$ based on our \SF\ data measured at the lowest pressure
($\sim$0.2~GPa). The constant level line $\kappa_b \simeq 0.85$,
i.e., the ``crossover line'', is depicted in
Fig.~\ref{fig:coherenceSF}. For pressures below 1.2~GPa the
crossover line exhibits a remarkable strong slope, not
expected in the naive picture of non-interacting electrons where
$T^* \propto t_b$. According to Fig.~\ref{fig:deconf} the
interstack transfer integral $t_b$ in \SF\ increases only by about 30\% 
when pressure increases by 4~GPa. Thus, one would expect the
crossover temperature of about 130~K at 4~GPa for non-interacting
electrons, in contrast to the experimental observation. This
provides evidence that electronic correlations play a decisive role
in the renormalization of the dimensional crossover in the Bechgaard
salts, leading to the very fast suppression of the 1D state in favor
of a high-dimensional metallic state. Unfortunately, the complete
theoretical description of this phenomenon is still not
elaborated.\cite{Giamarchi07}

The strong increase in the crossover temperature with pressure is
in agreement with the results of Moser {\it et al.} deduced from dc transport.\cite{Moser98}
The results of these earlier investigations are included in Fig.~\ref{fig:coherenceSF}
for comparison. In contrast to the almost linear crossover line based on the
dc transport results, our infrared spectroscopic measurements yield a sublinear 
behavior with a tendency of saturation towards higher pressures
(see Fig.~\ref{fig:coherenceSF}). This sublinear behavior could be
a consequence of the non-linear pressure dependence of $t_b$ which
shows the tendency for saturation (see Fig.~\ref{fig:deconf} and
discussion in Section~\ref{deconfinement}) together with
renormalization effects due to electronic interactions.
However, we cannot completely rule out a certain effect of the non-perfect hydrostaticity
in our measurements due to the solid pressure-transmitting medium, especially at high pressures;
this may contribute to the observed increase in the scattering rate of the Drude response, which
lowers the coherence parameter and shifts the dimensional crossover line down to
lower temperatures.

\begin{figure}
  \centerline{
  \includegraphics[width=0.95\columnwidth]{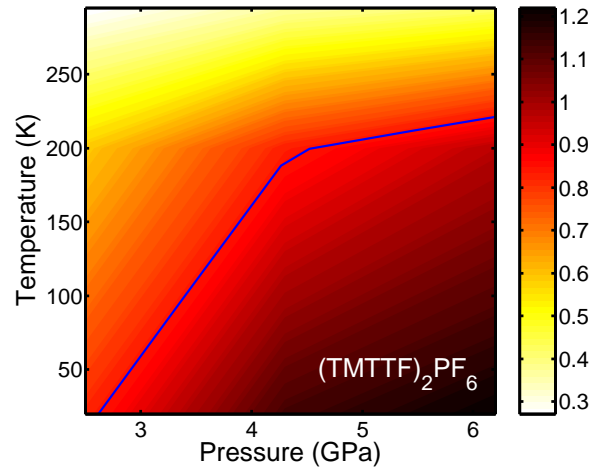}
  }\caption{(Color online) The coherence parameter of the inter\-stack charge transport 
  $\kappa_b$ as a function of temperature and pressure
  for \TF. The solid blue line corresponds to $\kappa_b = 0.85$.
  }\label{fig:coherenceTF}
\end{figure}

A corresponding temperature-pressure diagram of the coherence parameter $\kappa_b$ is 
depicted in Fig.~\ref{fig:coherenceTF} for \TF\ in the deconfined
state, i.e., for $P>2$~GPa. Due to the limited pressure range, the number of
experimental points amounts to three temperatures for each of the three
pressure values. According to Fig.~\ref{fig:coherenceTF} the coherence parameter
demonstrates a temperature and pressure dependence similar to that of the \SF\ salt. 
In Fig.~\ref{fig:coherenceTF} we include the dimensional crossover line following 
the critical value deduced for the \SF\ salt $\kappa_b = 0.85$. Also in the case of
\TF\ this crossover line exhibits a sublinear character and agrees rather well 
with the crossover line of the selenium analogue shifted by 3~GPa along the
pressure scale. Thus, the chemical pressure offset of 3~GPa
between the two studied salts in the generic phase diagram also holds for the
occurrence of the dimensional crossover.

\section{Summary}\label{Summary}

Our comprehensive investigations of the infrared response of
quasi-1D Bechgaard-Fabre salts as a function of pressure and
temperature demonstrate that the pressure-induced deconfinement
transition in the Mott insulator \TF\ occurs at
approximately 2~GPa. This critical pressure is basically temperature
independent and it is characterized by the rapid onset of the
interstack electronic transport (along the $b'$ direction). The size
of the Mott gap, $\Delta_\rho$, rapidly decreases as the
transition point is approached and stabilizes at a finite value upon
further pressure increase. The deconfinement occurs when
$\Delta_\rho \approx 2 t_b$. These findings are in accordance with
theoretical predictions and earlier experiments which utilized the effect of
chemical pressure.

Furthermore, the dimensional crossover in the studied salts occurs
when the coherence parameter of the interstack transport $\kappa_b$
exceeds a critical value of 0.85. The crossover
temperature increases rapidly with pressure, which evidences the
importance of electronic correlations for the dimensional
crossover. The line of the dimensional crossover is sublinear in the
temperature-pressure diagram and saturates at high pressures.

\begin{figure}
  \centerline{
  \includegraphics[width=1\columnwidth]{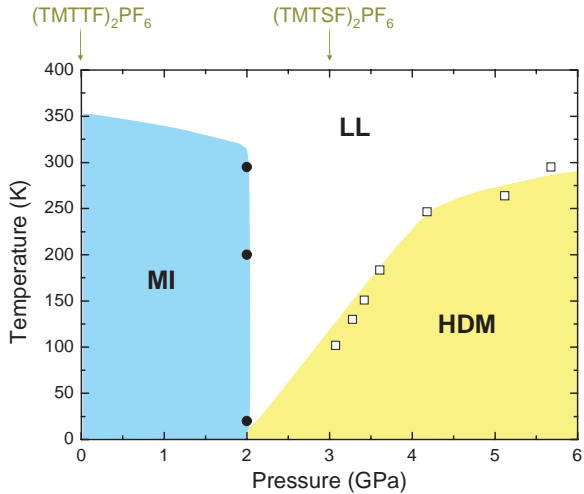}
  }\caption{(color online) Temperature-pressure phase diagram of the Bechgaard-Fabre
  salts under investigation. The symbols represent the phase boundaries determined in this work:
  The deconfinement transition (solid circles) from the Mott insulator
  (MI) into the 1D Luttinger liquid (LL), and the dimensional crossover (open squares)
  from the LL to the high-dimensional metallic state (HDM).
}\label{fig:summary}
\end{figure}

Remarkably, the pressure offset between the dimensional crossover lines for
\TF\ and \SF\ amounts to 3~GPa, i.e., the same as the corresponding offset between 
the values of the transverse hopping integral $t_b$. All the results obtained within 
this work can thus be included in a unified temperature-pressure phase diagram, which
is presented in Fig.~\ref{fig:summary}. The experimental points include the 
($P,T$) values of the deconfinement transition in \TF\ and the dimensional 
crossover line in \SF\ shifted by 3~GPa along the pressure axis. Based on all data 
we schematically depict colored regions corresponding to different phases, namely 
the Mott insulating, Luttinger liquid, and the high-dimensional metallic state.

\section{Acknowledgements}

We thank G. Untereiner for crystal growth. We
acknowledge the ESRF and ANKA facilities for the provision of
beamtime and we would like to thank B. Gasharova, Y.-L. Mathis, D.
Moss, and M. S\"upfle for help at the IR beamline. We also would
like to thank N. Drichko,  M. Dumm, S. Ebbinghaus, T. Giamarchi, E. Rose,
and K. Thirunavukkuarasu for fruitful discussions
and K. Syassen for providing valuable information about the
construction of the home-made infrared microscope. Financial
support by the DFG (Emmy Noether-program, SFB 484, DR228/27) is gratefully
acknowledged.


\begin{thebibliography}{52}
\bibitem[{com({\natexlab{a}})}]{comment0}
\bibinfo{note}{The crystals have triclinic symmetry. The $a$ axis is along the
  stacks formed by the planar organic molecules, $b^{\prime}$ denotes the
  projection of the $b$ axis perpendicular to $a$, and $c^*$ is normal to the
  $ab$ plane.}
\bibitem[{com({\natexlab{b}})}]{comment3}
\bibinfo{note}{At high pressure the emv coupling becomes stronger and the
  phonons become more asymmetric and broader. Thus the fit of the phonon
  contributions is less perfect, leading to the weak shoulder around 800 to
  1000~\icm\ in the spectra of \TF.}
\bibitem[{com({\natexlab{c}})}]{comment4}
\bibinfo{note}{DC experiments under pressure performed in the $b'$ direction
  confirm this behavior: Although the absolute value of the room-temperature
  resistivity decreases with pressure, $\rho(T)$ resembles an activating
  hopping transport at low-pressure. E. Rose, M. Dumm, and M. Dressel (to be
  published).}
\bibitem[{com({\natexlab{d}})}]{comment1}
\bibinfo{note}{Note that the spectral weights of the transverse optical
  conductivity shown in the insets of Fig.~\ref{fig:sigma-b} behave differently
  for both salts. Only the electronic hopping integrals calculated using
  pressure dependent unit cell parameters demonstrate agreement with the
  generic temperature-pressure phase diagram.\cite{Bourbonnais98}}.
\end{thebibliography}
\end{document}